\documentclass{PoS}

\usepackage{subfigure}
\usepackage{amsmath}

\title{New probes for bino dark matter with coannihilation at the LHC}

\ShortTitle{New probes for bino dark matter with coannihilation at the LHC}

\author{\speaker{H.~Otono}\\
        Research Centre for Advanced Particle Physics, Kyushu University, Fukuoka, Japan\\
        E-mail: \email{otono@phys.kyushu-u.ac.jp }}

\abstract{
It has been widely known that bino-like dark matter in supersymmetric theories suffers from over-production. 
The situation can be improved if the gluino or wino has a mass of $O(10)~\rm{GeV}$ heavier than the bino, 
sufficiently reducing the bino abundance through co-annihilation.
In this scenario, the gluino decays to the bino via squark exchange, and the wino decays to the bino via higgsino exchange.
In split SUSY models favoured after the Higgs discovery, 
the intermediate particles in these decays would be much heavier than gauginos, suppressing the decay of the gluino and wino.
This, in addition to the small mass differences, results in long lifetimes for the gluino and wino. 
We show that searches performed at the LHC for long-lived particles with displaced vertices offer a powerful method to test this scenario.
}

\FullConference{The European Physical Society Conference on High Energy Physics\\
		22--29 July 2015\\
		Vienna, Austria}

\begin{document}

\section{Introduction}
\label{Introduction}

Although there are strong evidences for dark matter (DM), we do not have any DM candidates in the Standard Model.
The lightest SUSY particle (LSP) with a mass of $O(0.1-1)~\rm{TeV}$ is a candidate for the DM.
However, according to quite a few experimental implications, e.g., the Higgs mass of $125~\rm{GeV}$ \cite{Aad:2015zhl},
squarks would be much heavier than the LSP in the Minimal SUSY Standard Model, namely more than $O(10)~\rm{TeV}$. 
Such a mass spectrum is realized by an anomaly-mediated SUSY breaking (AMSB) scenario \cite{Giudice:1998xp, Randall:1998uk}, called mini-split SUSY \cite{Mini-split}.
Considering that the higgsino mass parameter $\mu$ is usually expected to be of the same order as the squark masses,
the LSP should be either pure wino or pure bino.
The AMSB models generally give rise to the pure wino LSP.
The pure bino LSP is also possible due to the higgsino threshold correction \cite{Giudice:1998xp, Pierce:1996zz} and 
the effect of extra particles \cite{Pomarol}.

The abundance of the LSP produced in the early Universe should be decreased to the level of the observed relic abundance,
which depends on the self-annihilation cross sections of the LSP.
However, the presence of the another gaugino, i.e., the gluino or wino, 
as the next lightest SUSY particle (NLSP) with a mass difference of $O(10)~\rm{GeV}$ relative to the bino,
allows for higher bino masses due to co-annihilation in the chemical equilibrium \cite{Griest:1990kh}.

These NLSPs becomes long-lived due to the intermediate heavy particles in the mini-split SUSY scenario 
-- the squarks for a gluino NLSP and the higgsino for a wino NLSP.
The phase-space for the decays of the NLSPs is also suppressed due to the small mass difference from the bino LSP. 
As a result, the emanation from the NLSPs produced at the LHC would leave the distinct signature inside the detectors in the form of a displaced vertex (DV) 
from the collision point.
The invariant mass of the charged tracks from the DV should be $O(10)~\rm{GeV}$.
The detailed detectability at the LHC is discussed in Refs \cite{bino-gluino, bino-wino}.

\section{Gaugino co-annihilations}\label{sec:co-anni}

\begin{figure}
    \subfigure[]{%
        \includegraphics*[bb=0 0 357 277, width=0.5\columnwidth]{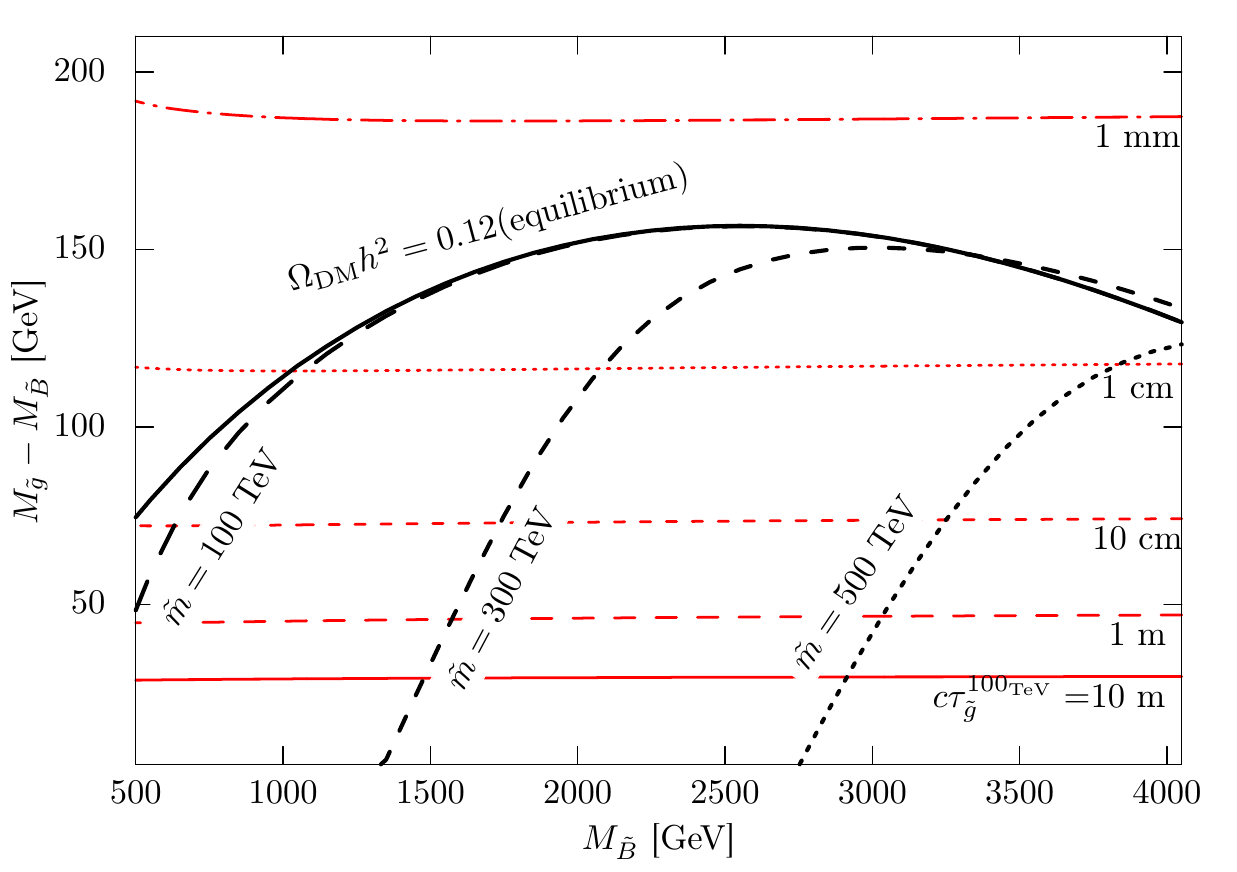}}%
		\label{fig:wwb}
    \subfigure[]{%
        \includegraphics[bb=0 0 360 252, width=0.5\columnwidth]{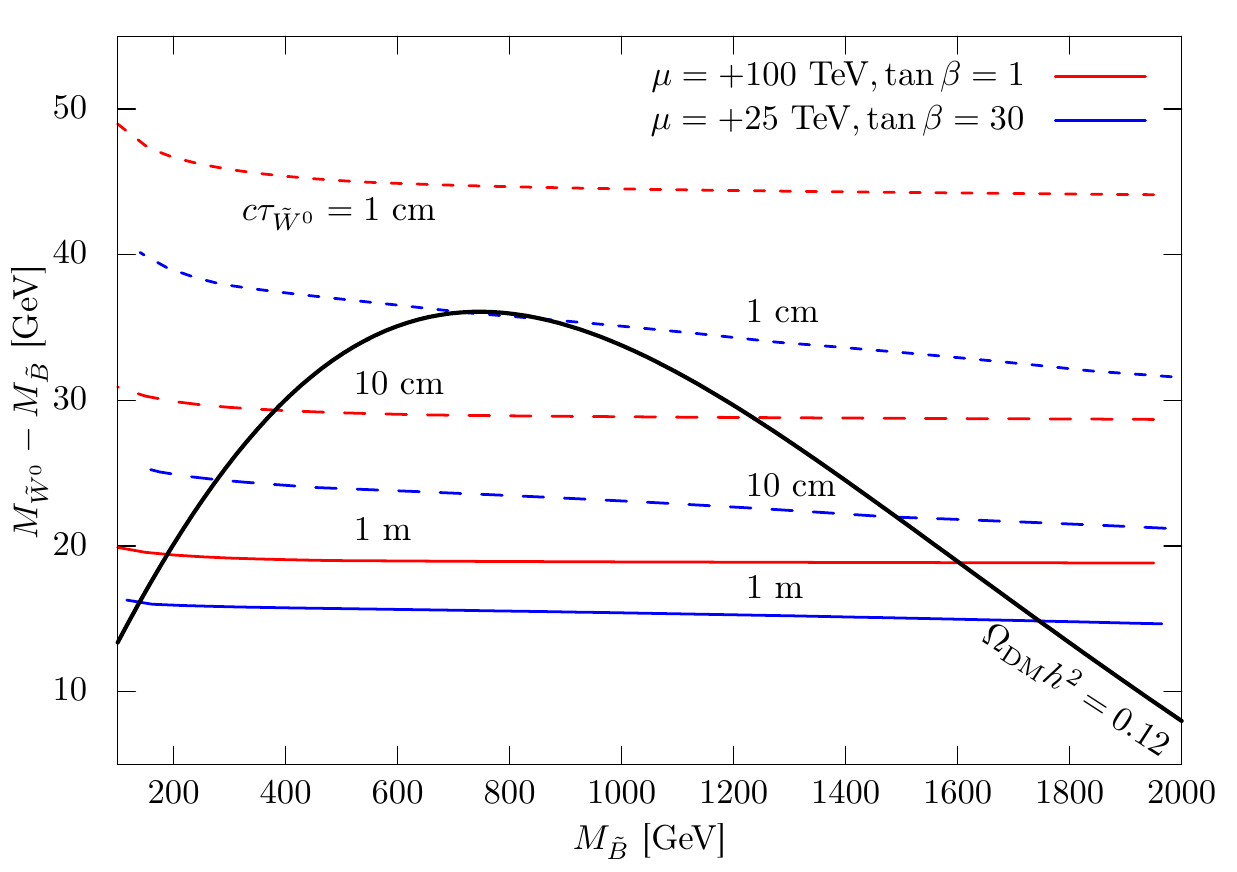}}%
		\label{fig:gwb}
    \caption{
	    Mass difference between the NLSP (the gluino or wino) and the bino LSP allowed by the observed DM density as a function of bino mass.
	    The co-annihilation scenario fails in the region above the black lines. 
		(a) Bino-gluino co-annihilation.
  		The colored lines show the points in this parameter space for which the proper flight length of the gluino ($c\tau_{\widetilde{g}}$) with the squark mass 
  		$\widetilde{m}=100~\rm{TeV}$ has the indicated value. 
		(b) Bino-wino co-annihilation.
 		The colored lines show the proper flight lengths of the neutral wino ($c\tau_{\widetilde{W^0}}$) with two sets of $\mu$ and $\rm{tan} \beta$.
    }
    \label{fig:massdiff}
\end{figure}

If mass difference between the bino LSP and the gluino NLSP, $\Delta M_{\widetilde{g}-\widetilde{B}}$, is small,
a bino can be converted to a gluino in the thermal bath, significantly reducing the bino abundance due to the large cross section of the gluino self-annihilation.
In order to efficiently realize this gluino-bino co-annihilation, 
the transition from bino to gluino should be faster than the Hubble expansion rate.
Since the transition is governed by the squark mass, $\widetilde{m}$, and 
the freeze-out temperature is determined by the bino mass, $M_{\widetilde{B}}$,
the following constraint can be obtained: 
\begin{equation*}
\widetilde{m} < 250~{\rm{TeV}} \times (\frac{M_{\widetilde{B}}}{1~\rm{TeV}}) ^ \frac{3}{4}.
\end{equation*}
Figure \ref{fig:massdiff} shows $\Delta M_{\widetilde{g}-\widetilde{B}}$ of less than around $100~\rm{GeV}$ is allowed from the observed DM density.
A $\widetilde{m}$ value of $O(100)~\rm{TeV}$ is possible, but then $\Delta M_{\widetilde{g}-\widetilde{B}}$ should be lighter than $100~\rm{GeV}$. 
The gluino NLSP decays to the bino LSP with emission of two quarks: $\widetilde{g} \rightarrow \widetilde{B}qq$. 
The lifetime of the gluino NLSP, $\tau_{\widetilde{g}}$, is suppressed by both $\Delta M_{\widetilde{g}-\widetilde{B}}$ and $\widetilde{m}$ as follows:
\begin{equation*}
c\tau_{\widetilde{g}} = O(1) \times (\frac{\Delta M_{\widetilde{g}-\widetilde{B}}}{100~\rm{GeV}})^{{-5}}(\frac{\widetilde{m}}{100~\rm{TeV}})^{4}~\rm{cm}.
\end{equation*}
For the allowed $\Delta M_{\widetilde{g}-\widetilde{B}}$ region, the gluino NLSP produced at the LHC has $c\tau_{\widetilde{g}}$ of more than $O(1)~\rm{mm}$ from the collision point.

The DM particle might be the bino LSP with the compressed winos giving rise to the co-annihilation.
The transition from a bino to winos in the thermal bath is mediated by a higgsino. 
In the mini-split SUSY, this process is so rapid compared to the Hubble expansion.
Figure \ref{fig:massdiff} shows the allowed mass difference between the bino and neutral wino, 
$\Delta M_{\widetilde{W^0}-\widetilde{B}}$, which is less than around $30~\rm{GeV}$.
Due to the heavy higgsino and the small $\Delta M_{\widetilde{W^0}-\widetilde{B}}$, the neutral wino has a long lifetime.  
Note that the charged wino promptly decays to the bino, not having significant proper flight length.
Assuming $|\mu| > O(10)~\rm{TeV}$ and $\tan \beta = O(1)$ in the mini-split SUSY,
the dominant process for the decay of the neutral wino is $\widetilde{W^0} \rightarrow \widetilde{B} h$.
The detailed calculation is shown in Ref \cite{bino-wino}.
In this study, we mainly focus on this Higgs boson mediated decay.

\section{LHC prospects}\label{sec:pros}

So far, various long-lived particle searches have been performed at the LHC.
In particular, the ATLAS collaboration has searched for long-lived gluino, making full use of the inner detectors \cite{ATLAS_0,ATLAS_1,ATLAS_2, ATLAS_3},
forcusing on DVs with a invariant mass of more than $O(100)~\rm{GeV}$.
From these results, we evaluated the sensitivities for the scenario proposed in this paper, which predicts DVs with a lower invariant mass.
For the signal generation, we used the program packages {\sc Herwig~6} \cite{Herwig} and {\sc AcerDET} \cite{ACER} for the long-lived gluino, and  
{\sc Madgraph~5} \cite{Alwall:2011uj}, {\sc Pythia~6} \cite{Sjostrand:2006za}, and {\sc Delphes3} \cite{deFavereau:2013fsa} for the long-lived neutral wino. 
{\sc Prospino2} \cite{Beenakker:1996ed} is used for the cross sections of the SUSY particles. 

\begin{figure}[htbp]
  \begin{center}
  \includegraphics[bb=0 0 355 277, width=100mm]{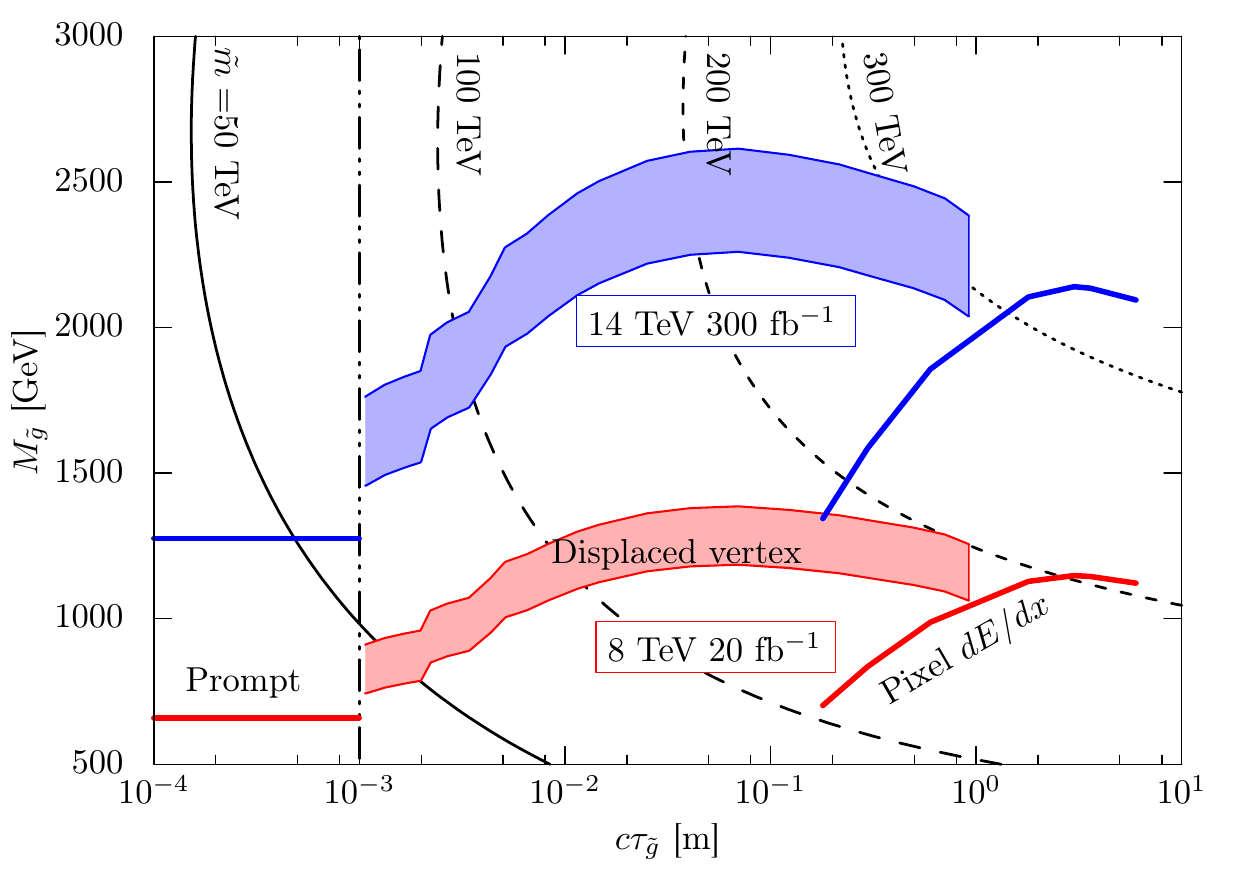}
  \end{center}
  \caption{Sensitivity for the gluino-bino co-annihilation (the blue and red bands) with a gluino-bino mass difference of $100~\rm{GeV}$.
  The prompt-decay gluino searches \cite{Prompt_1, Prompt_2} and the Pixel $dE/dx$ search \cite{Pixel}  are also shown.
  The black lines describe the relation among the squark mass, the gluino mass and $c\tau_{\widetilde{g}}$, assuming the $\Delta M_{\widetilde{g}-\widetilde{B}}$ as
  the abundance of the bino LSP would be equal to the observed DM density.}
  \label{fig:pros_ggb}
\end{figure}

With regard to triggering the events, the truth-level missing transverse energy, $E^{\rm{miss}}_{\rm{T}}$, was employed. 
For the centre of mass energy of $8~\rm{TeV}$ and $14~\rm{TeV}$, the trigger requirements
$E^{\rm{miss}}_{\rm{T}} > 100~\rm{GeV}$ and $200~\rm{GeV}$ are applied, respectively.
Figure \ref{fig:pros_ggb} shows the sensitivity for the gluino-bino co-annihilation.
The upper and the lower bounds of each band assume that the reconstruction efficiency for the DV is $100\%$ and $20\%$, respectively,
which is extracted from the results in Ref \cite{ATLAS_2}.
The sensitivity was calculated so as to obtain three signal events. 
With an expected background of 0.5 events, the bands corresponded to a $3\sigma$ evidence sensitivity.
Note that the expected background level was 0.01 events for the ATLAS search with $E^{\rm{miss}}_{\rm{T}}$ trigger in Ref \cite{ATLAS_3}.
Consequently, a $300~\rm{fb}^{-1}$ dataset collected with a collision energy of $14~\rm{TeV}$ at the LHC would be sensitive to production of a long-lived gluino with 
a mass of more than $2000~\rm{GeV}$ as shown in Figure \ref{fig:pros_ggb}. 
The actual search with the $20~\rm{fb}^{-1}$ dataset collected with the $8~\rm{TeV}$ collision has not been performed yet.

\begin{figure}
    \subfigure[]{%
        \includegraphics*[bb=0 0 356 252, width=0.5\columnwidth]{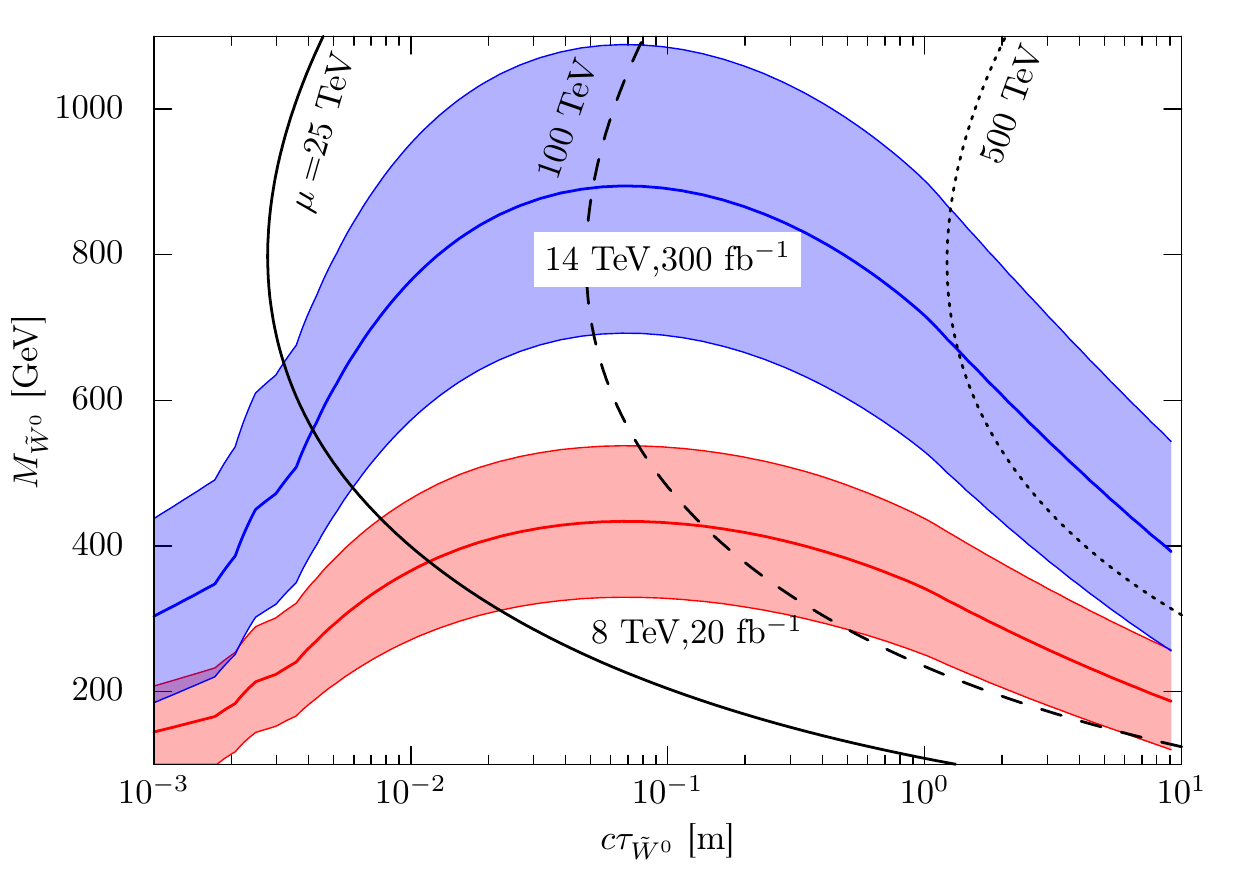}}%
		\label{fig:wwb}
    \subfigure[]{%
        \includegraphics[bb=0 0 356 252, width=0.5\columnwidth]{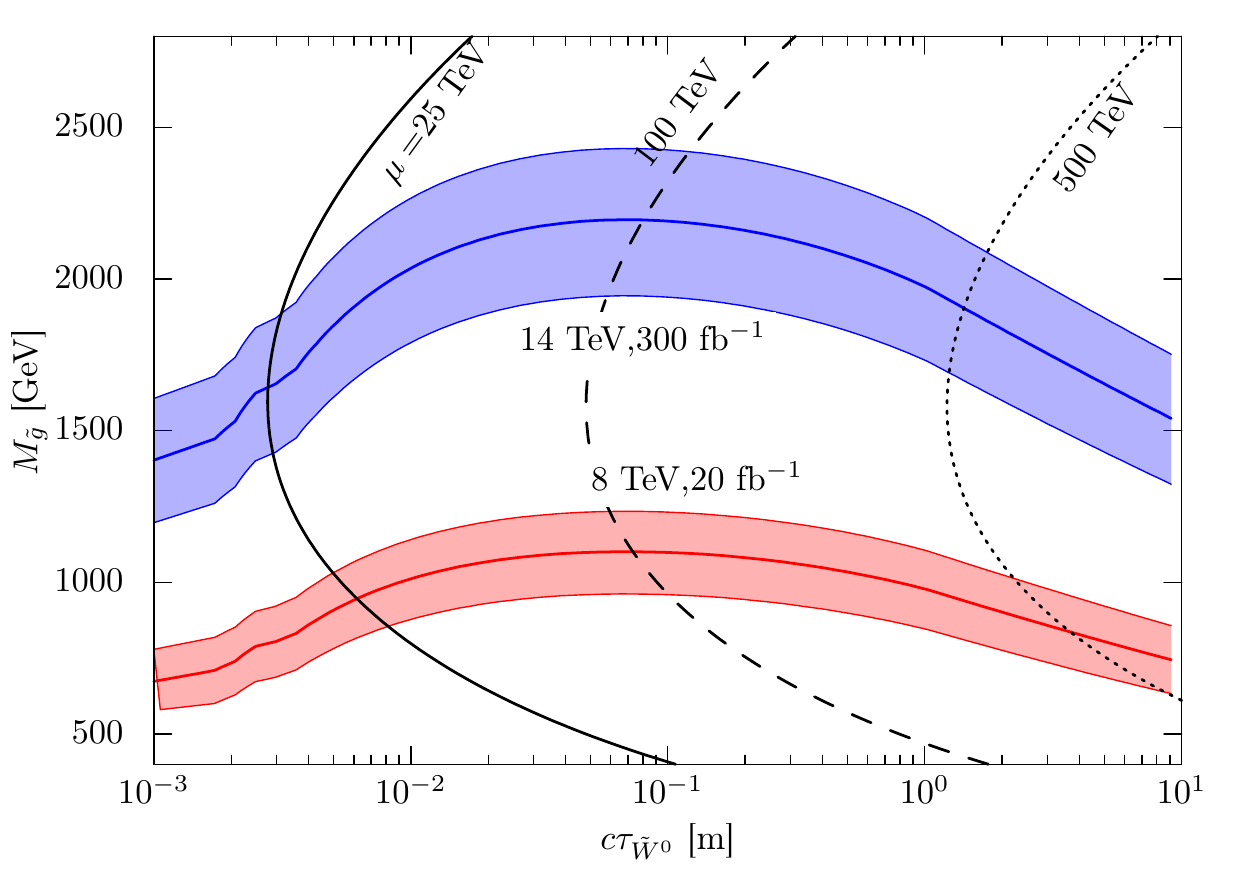}}%
		\label{fig:gwb}
    \caption{Sensitivity for the wino-bino co-annihilation, assuming the $\Delta M_{\widetilde{W^0}-\widetilde{B}}$ 
    as the abundance of the bino LSP would be equal to the observed DM density.
    Black lines describe the relation among $\mu$, the neutral wino mass and $c\tau_{\widetilde{W^0}}$ under $\rm{tan}\beta=2$.
    (a) Neutral and charged wino pair production,
    (b) Gluino pair production with $M_{\widetilde{g}}=2\times M_{\widetilde{W^0}}$,
	assuming the branching fraction from a gluino to a neutral wino is $30\%$.
    }
    \label{fig:wb}
\end{figure}

For the wino-bino co-annihilation scenario, the trigger efficiency could be estimated as we did for the gluino-bino co-annhilation.
However there is no reference for the reconstruction efficiency for a DV with a mass difference of $O(10)~\rm{GeV}$.
Therefore, the following conditions, which were required in the ATLAS search \cite{ATLAS_0,ATLAS_1,ATLAS_2, ATLAS_3}, 
were applied to obtain the reconstruction efficiency
for the signal sample: 
the DV was required to contain more than four tracks, each with a transverse momentum of $P_{\rm{T}} > 1~\rm{GeV}$, 
and the invariant mass of the DV tracks must be more than $10~\rm{GeV}$.
We considered two processes for the wino-bino co-annihilation.
Figure \ref{fig:wb} shows the sensitivities for neutral and charged wino pair production, and gluino pair production 
where a gluino sometimes decays to a bino via a neutral wino.
We assumed $M_{\widetilde{g}}=2\times M_{\widetilde{W^0}}$ and $30\%$ of branching fraction for $\widetilde{g}\rightarrow \widetilde{W^0} \rightarrow \widetilde{B}$.
The line at the center of each band in Figure \ref{fig:wb} was obtained from the estimated DV reconstruction efficiency with $\Delta M_{\widetilde{W^0}-\widetilde{B}} = 30~\rm{GeV}$.
The upper (lower) bound shows the case where the efficiency is three times better (worse) than expectation.
The long-lived neutral wino emits a virtual Higgs boson, which decays mainly to two long-lived $b$-hadrons,
giving even more characteristic signature -- two displaced vertices originating from the long-lived neutral wino decay vertex.
However, we have not considered the reconstruction of this specific signature. 
Nonetheless, we found that, even when not attempting to reconstruct the $b$-hadron vertices separately, significant sensitivity to this scenario was obtained.

\section{Conclusion}\label{sec:conclusion}

The observed DM density can be explained by a bino LSP with a mass of $O(0.1-1)~\rm{TeV}$ with co-annihilation with the gluino or wino NLSP.
Several searches put lower bounds of $O(10)~\rm{TeV}$ on the squark masses.
Such a mass spectrum can be realised in the mini-split SUSY scenario, which naturally makes the gluino or wino NLSP long-lived.
Based on the ATLAS search for a displaced vertex from the collision point,
we found that the $14~\rm{TeV}$ LHC would have a great capability to discover the long-lived NLSPs.

\section*{Acknowledgement}\label{sec:conclusion}
This work was supported by JSPS KAKENHI Grant Number 15K17653.

\end{document}